\documentclass{emulateapj}

\shorttitle{Consequences of GW recoil for unified models of AGN}
\shortauthors{Komossa \& Merritt}

\def\approxlt{\mathrel{\hbox{\rlap{\lower.55ex \hbox {$\sim$}}
        \kern-.3em \raise.4ex \hbox{$<$}}}}
\def\approxgt{\mathrel{\hbox{\rlap{\lower.55ex \hbox {$\sim$}}
        \kern-.3em \raise.4ex \hbox{$>$}}}}
\def\mh{M_{\rm BH}}
\def\mgal{M_{\rm gal}}
\def\msun{M_\odot}
\def\vk{v_{\rm k}}

\begin{document}

\title{Gravitational wave recoil oscillations of black holes: Implications for unified models of Active Galactic Nuclei}

\author{S. Komossa}
\affil{Max-Planck-Institut f\"ur extraterrestrische Physik,
Postfach 1312, 85741 Garching, Germany; skomossa@mpe.mpg.de}
\author{David Merritt}
\affil{Center for Computational Relativity and Gravitation and
Department of Physics, Rochester Institute of Technology,
Rochester, NY 14623, USA; merritt@astro.rit.edu}

\begin{abstract}
We consider the consequences of gravitational wave recoil
for unified models of active galactic nuclei (AGNs).
Spatial oscillations of supermassive black holes (SMBHs) around the
cores of galaxies following gravitational wave (GW) recoil imply that the SMBHs spend a 
significant fraction of time off-nucleus, at scales beyond that of the molecular
obscuring torus.  
Assuming reasonable
distributions of recoil velocities, we 
compute the off-core timescale of (intrinsically type-2)
quasars.
We find that roughly one-half of
major mergers result in a SMBH being displaced beyond the torus
for a time of $10^{7.5}$ yr or more, comparable to quasar activity timescales. 
Since {\em major}
mergers are most strongly affected by GW recoil,
our results imply a deficiency of type 2 quasars 
in comparison to Seyfert 2 galaxies.
Other
consequences of the recoil oscillations for the observable properties
of AGNs are also discussed.
\end{abstract}


\section{Introduction}

Gravitational waves emitted anisotropically during gravitational collapse
carry away linear momentum. As a result 
the center of mass of the collapsing object recoils
(Peres 1962; Bekenstein 1973).
Configurations of coalescing spinning black holes can result in  
recoil velocities of hundreds to thousands of km\,s$^{-1}$
(e.g., Campanelli et al. 2007; Gonz{\'a}lez et al. 2007;
Herrmann et al. 2007; Pollney et al. 2007; Tichy \& Marronetti 2007;
Br\"ugmann et al. 2008; Dain et al. 2008), 
scaling to a maximum
of $\sim 4000$ km\,s$^{-1}$ (Campanelli et al. 2007; Baker et al. 2008)
for maximally spinning equal-mass binaries with anti-aligned spins
in the orbital plane, and as large as $\sim 10^{4}$ km\,s$^{-1}$ in hyperbolic encounters
(Healy et al. 2008).
Kicks large enough to remove SMBHs from galaxies 
have potentially far-reaching consequences for SMBH and galaxy assembly,
and predict interstellar and intergalactic quasars 
(e.g.,  Madau et al. 2004; Merritt et al. 2004; Madau \& Quataert 2004;
Haiman 2004;
Boylan-Kolchin et al. 2004;
Libeskind et al. 2006; Loeb 2007; Schnittman 2007; Volonteri 2007;
Gualandris \& Merritt 2008; Volonteri \& Madau 2008, Kornreich \& Lovelace 2008).

Komossa et al. (2008) reported the detection of a 
recoil candidate with a projected kick velocity
of 2650 km s$^{-1}$. The quasar SDSSJ092712.65+294344.0
shows three unusual emission-line systems,
including a kinematically offset broad-line region (BLR) 
and a system of atypically narrow emission lines which
lack the usual ionization stratification -- two key signatures of kicks. 
Apart from spectroscopic signatures (Merritt et al. 2006; Bonning et al. 2007),
offset quasars could be detected  for instance  
by their temporarily flaring accretion disks (Shields \& Bonning 2008;
Lippai et al. 2008; Schnittman \& Krolik 2008), 
by tidal disruption flares from the bound
(and unbound) population of stars and
episodic fuelling from
stellar mass loss 
(Komossa \& Merritt 2008), and via the bound compact
star cluster itself (Merritt et al. 2008, O'Leary \& Loeb 2008). 
One key consequence of gravitational wave recoil is long-lasting oscillations 
of the SMBH about the galaxy core, 
implying that SMBHs may spend as long as 10$^{6-9}$ yrs off-nucleus  
with an amplitude of parsecs or kiloparsecs, depending 
on kick velocity and galaxy structural parameters 
(Merritt et al. 2004; Madau \& Quataert 2004; Gualandris \& Merritt 2008;
MQ04 and GM08 hereafter).
In this Letter, we consider consequences of these ``recoil oscillations''
for unified models of active galactic nuclei (AGNs).  

According to unified models, AGNs are intrinsically similar, 
but their appearance depends strongly on the line of sight of the
observer toward the ``central engine'' (reviews by Antonucci 1993;
Elitzur 2007). Along certain sightlines, a dusty torus
consisting of molecular clouds blocks the observer's view, hiding
some  core components, especially the BLR, which therefore
is only directly visible in ``type 1'' AGNs.   
The unified model has been very successful in
explaining observed properties of AGNs including 
the presence of hidden BLRs detected in polarized light (e.g.,  Antonucci \& Miller 1985;
Zakamska et al. 2005),  
and has been corroborated by recent imaging and spectroscopy of the
torus (e.g., Jaffe et al. 2004; Siebenmorgen et al. 2005). 

If SMBHs and the BLRs bound to them
spend a significant time displaced from the nucleus, this will have profound
consequences for obscuration-based unified models of AGNs. 
This statement holds true whether the obscuration originates in
parsec-scale molecular tori (Antonucci 1993), or
compact star-forming regions  (Levenson et al. 2001),
or is due to absorption associated with the host galaxy itself,
which may work on typical scales of $\sim$100 pc
(Maiolino \& Rieke 1995).
Here we argue that
a significant fraction of the quasar population is expected
to be in a regime such that the SMBH and the BLR bound to it 
is displaced beyond the obscuring region,
implying a deficiency of type 2 (obscured) AGNs among quasars.  

\section{Relevant spatial scales in the AGN core}

We first compare
three characteristic spatial scales: the kick radius, the BLR size, and the torus size.
After the kick, matter remains bound to the SMBH within a region 
whose radius $r_{\rm k}$ is given by  
\begin{equation}
r_{\rm k} = {G\mh\over{v_{\rm k}^2}}
\approx 0.43
 \left({\mh\over{10^8 M_\odot}}\right)
\left({v_{\rm k}\over{10^3 {\rm km\ s}^{-1}}}\right)^{-2} ~{\rm pc}
\end{equation}
where $v_{\rm k}$ is the kick velocity (e.g., Merritt et al. 2006). 
The size $r_{\rm BLR}$ of the BLR of AGNs has been determined from 
reverberation mapping (Peterson 2007), 
and scales with AGN luminosity as
\begin{equation}
r_{\rm BLR} = 0.1 
\left({{\lambda\,L_{\lambda}{\rm(5100\AA)}}
\over{10^{45}{\rm erg\ s}^{-1}}}\right)^{0.69} ~{\rm pc}
\end{equation}
(Kaspi et al. 2005)
where the luminosity at 5100\AA, 
$\lambda L_{\lambda}{\rm(5100\AA)} \simeq 0.1\,L$ and $L$ is
the AGN bolometric luminosity.
The size of the molecular torus is still poorly constrained
from observations. Recent measurements of dusty 
gas in the Seyfert galaxies NGC\,1068 and Circinus 
suggest an extent of 2-3 pc (Jaffe et al. 2004;
Davies et al. 2007; Tristram et al. 2007).
It is reasonable to assume that the {\em inner} edge of the torus, $r_{\rm tor,in}$, 
is beyond the dust sublimation radius (e.g., Netzer 1990; Nenkova et al. 2008)
which is given by 
\begin{equation}
r_{\rm tor, in} \approxgt 0.4 \left({L\over{10^{45}{\rm erg\ 
s}^{-1}}}\right)^{1/2}
\left({1500{\rm K}\over{T_{\rm sub}}}\right)^{2.6}~ {\rm pc}
\label{eq:rtorus}
\end{equation}
with an outer radius likely not much larger than $\sim 20\,r_{\rm tor, in}$ (Elitzur 2007),
and where $T_{\rm sub}$ is the dust sublimation temperature.
Comparison of these three relations shows that a large fraction of the BLR 
remains bound to the recoiling hole, while
structures of the size of the torus or larger will typically be left behind. 
Oscillation amplitudes of $\approxgt$ 10-20 pc will therefore move the SMBH
beyond the torus scale, except for the highest SMBH masses, where part of
the torus will remain bound.   

\section{Oscillation amplitudes and oscillation durations in quasars}

We base our discussion on the $N$-body simulations 
of GM08, 
who computed SMBH trajectories after GW recoil.
Key model parameters are the SMBH mass $\mh$
and galaxy mass $M_{\rm gal} = 10^{3} \mh$,
the kick velocity $v_{\rm k}$, and the galaxy structural
parameters.
We concentrate here on the mass range of SMBHs
that is typical for the bulk of quasars 
($M_{\rm{BH}} \approx 10^{8}- {\rm few}\,10^{9}\msun$).
Massive inactive elliptical galaxies and host galaxies
of luminous quasars typically have effective radii $R_{\rm e}$ 
on the order of a few to $\sim10$ kpc. 
Here, we adopt the relation between $R_{\rm e}$ and $\mgal$
defined by the nine luminous ($-22\lesssim M_V\lesssim -24$)
quasar host galaxies in the sample of 
Wolf \& Sheinis (2008),
i.e. $\log (R_{\rm e}/{\rm kpc}) \approx -5.61 +  0.55\log(\mgal/\msun)$.
The models of GM08 on which we base our discussion had post-kick
core radii $r_{\rm c}$ consistent with the range defined by the brightest
E-galaxies with resolved cores (Ferrarese et al. 2006),
i.e. $0.01\lesssim r_{\rm c}/R_{\rm e}\lesssim 0.03$.
We  focus here on the spherical ``A1'' galaxy models from GM08;
scaling of the $N$-body results to physical units was done
following their equation~(4).
Dark matter halos were ignored.

Figure~1a shows trajectories of kicked SMBHs scaled to a galaxy
with SMBH mass $5\times 10^{8}\msun$.
Kick velocities were $v_{\rm k}/v_{\rm esc}$ = 0.3,0.5,0.7,0.9,
corresponding to $v_{\rm k}=(360, 590, 830, 1070)$ km s$^{-1}$
for a central escape velocity $v_{\rm esc}$ of $1185$ km s$^{-1}$ of our galaxy model.
As discussed by GM08, SMBH oscillations continue well beyond the
time (``Phase I'') that would be predicted by applying Chandrasekhar's dynamical
friction formula assuming a fixed galaxy core 
(e.g. MQ04; Blecha \& Loeb 2008).
The time of onset of these long-term, or ``Phase II'', oscillations
is indicated by the open circles in Figure~1a.
Including the effect of the Phase II oscillations, Figure 1a shows
that the SMBH's motion 
persists for more than $\sim 10^7$ yr if $\vk\gtrsim 450$ km s$^{-1}$.

\section{Comparison with observations}

\subsection{Unified models and type1/type2 fractions}

About 70\% of the nearby Seyfert galaxies are type 2 
(e.g., Schmitt et al. 2001),
i.e. they lack a BLR in their (unpolarized) optical spectra. 
The fraction of the high-luminosity equivalents, type 2 quasars,
is less well known and still subject to a number of selection
effects (e.g. Halpern \& Moran 1998; Reyes et al. 2008),
even though studies generally indicate a deficiency of type 2 quasars. 
While in past studies type 2 quasars were observationally very rare or
absent, significant numbers have been found in recent X-ray and 
optical surveys (e.g., Norman et al. 2002; Zakamska et al. 2005; Brusa et al. 2007).
There is a systematic trend such that the fraction of type 2 sources, or
equivalently the amount of X-ray absorption, decreases with increasing
source luminosity (e.g., Simpson 2005; Barger et al. 2005;
Reyes et al. 2008; Hasinger 2008),
while the simplest possible version of the unified model would imply  
a constant type1/type2 ratio. 
Models have been proposed which account for this luminosity dependence,
e.g. by invoking changes in the properties
of the obscurer as $L$ increases (e.g., Nenkova et al. 2008; Ballantyne 2008). 
Recoil oscillations inevitably affect the numbers of obscured versus unobscured sources
and therefore have potentially profound implications for unified models.
How do they affect the ratio of type1/type2 quasars in comparison
to the number of type1/type2 Seyfert galaxies ?
There is increasing evidence that quasar activity is powered by major mergers
while Seyfert activity
may have other triggers including bars, minor mergers, or random accretion of molecular
clouds (e.g., Sanders et al. 1988; Urrutia et al. 2008; Hopkins et al. 2008; Hasinger 
2008, and references therein). 
Recoil oscillations have highest amplitudes in major mergers, and we may therefore
expect that the frequency and properties of (type 2) quasars are strongly affected by
recoil oscillations.
Can this explain the relative scarcity of type 2 quasars in comparison with type 2
Seyferts, and the more general trend that type2-ness (X-ray absorption) 
decreases with luminosity? 

\subsection{Rate estimates}

Since a large fraction of quasars are believed to be triggered by 
major mergers,
in order to estimate the fraction of quasars that occur at or above
a given kick velocity,
we carried out rate estimates relevant for major mergers with random spin
distributions (Campanelli et al. 2007; Schnittman \& Buanano 2007; Baker et al. 2008;
note that the actual kick velocities could be smaller in gas-rich systems if the mechanism
discussed by Bogdanovi{\'c} et al. 2007 is at work, but see Sect. 3 of Schnittman \& Krolik 2008).
We first updated these previous estimates, based on the recent kick formula of
Baker et al. (2008; essentially identical
results were obtained using the Lousto \& Zlochower 2008
version of the kick formula), and 
assuming random orientations of the spin vectors of both SMBHs
and a distribution
of SMBH mass ratios in the range $0.3\le q\le 1.0$ where
$q\equiv m_2/m_1, m_2\le m_1$, relevant for major mergers.
SMBH spins were drawn from a distribution such that
$a_1\le a\le a_2$ with $a\equiv S/m^2$ the dimensionless spin,
with $a_1=0.5$, $a_2=0.9$.
The distributions of $m$ and $a$ were assumed to be uniform
in the logarithm between these limits.
For this base model, we find that about 50\% of major mergers have kick
velocities above 500 km/s.   
At or above these kick velocities, {\em total} oscillation timescales
start to be on the order of 
quasar lifetimes, which are about 10$^{7-8}$ yr (e.g., Kauffmann \& Haehnelt 2000;
Yu \& Tremaine 2002; Hopkins et al. 2005; see review by Martini 2004). 

How much of the total quasar population is ultimately
affected by recoil oscillations then depends on (1) the time
$t_{\rm vis}$ the kicked SMBH+BLR spends {\em beyond}
the obscuring torus (i.e., at distances greater than
$r_{\rm torus}\approx r_{\rm tor,out}$
from the galaxy center (eqn.~\ref{eq:rtorus}))
and therefore appears as ``type 1'' rather than ``type 2'',
in comparison to (2) the total quasar lifetime.
We computed $t_{\rm vis}$ on a grid in $r_{\rm torus}$
for each of the nine $N$-body trajectories in GM08
($0.1\le\vk/v_{\rm esc}\le 0.9$).
Given arbitrary values of $\vk$, $\mgal$ and $r_{\rm torus}$, the
value of $t_{\rm vis}$ was then computed via interpolation
between the nine discrete kick velocities of GM08.

Figure~1b shows, for our base model, the fraction $f$ of kicked SMBHs that
spend more than $3\times 10^7$ yr at $r> r_{\rm min}$, for a range
of ($\mgal, r_{\rm min}$) values.
For $r_{\rm min} \approx r_{\rm tor,out}$, this fraction is
$\sim 0.5\pm 0.1$ with a weak dependence on galaxy mass.
We conclude that {\it a significant fraction of mergers would
result in recoiling SMBHs that spend of order a quasar lifetime
above the obscuring torus} and would therefore appear as type 1 quasars
even if ``intrinsically'' type 2. 
We examined the robustness of these results under different assumptions
about the pre-recoil distributions of spins and masses.
Maximally spinning, equal-mass 
SMBH mergers affect a fraction of up to $f \sim0.75$ of the population
(same assumptions as above), while major mergers with intermediate SMBH spins of $a=0.3$ imply $f \sim 0.05-0.15$.   
{\footnote{We note that galaxy triaxiality, not
included in the current models, might extend the oscillation timescales
by a factor of several (e.g. Vicari et al. 2007; see also MQ04 and Merritt et al. 2004);
inhomogeneities of the host galaxy after merging would likely
have the same effect.
}}

If we assume that the type1/type2 fraction of quasars is 
intrinsically (i.e., in the absence of recoils) the same as in Seyferts 
($\sim$70\% type 2, $\sim$30\% type 1), 
we predict that a fraction $\sim50\%$ of
these type2s will appear instead as type 1s 
at any given time. 
Several factors affect these estimates: idealizations in the
galaxy models used in the 
$N$-body simulations, uncertainties in the distribution
of masses and spins
as discussed above, uncertainties especially in the thickness of the torus
in the most massive galaxies, and various selection
effects in the measurements of type1/type2 ratios in 
dependence of SMBH mass. 
We also recall that if equations (1) and (3) strictly hold, in the most
massive quasars a fraction of the torus will remain bound to the
recoiling SMBH
so absorption/extinction would not fall to zero. 
Finally, we note that if the quasars' {\em total} lifetime is actually
composed of several shorter merger episodes on the order of 10$^{6}$ yr
each, even recoil oscillations with velocities as small as $\sim 200$ km s$^{-1}$
would affect a large fraction of the quasar population.

\subsection{Implications}

So far, we have distinguished between Seyfert galaxies (low-mass SMBHs)
and quasars (high-mass SMBHs). Can we also
reproduce the observed trend (Sect. 4.1) that obscuration fraction 
decreases with quasar luminosity (i.e., mass)?
In our base model, the dependence of mean oscillation
timescale on galaxy mass is weak.
However, 
since the likelihood of a major merger (as opposed to other types of fuelling)
is believed to increase strongly with AGN luminosity (galaxy mass), 
the observed trend should arise naturally, since the most luminous AGNs 
are increasingly likely to be triggered by major mergers.

If a fraction of all quasars is recoiling at any given time, 
we should see the corresponding BLR emission-line velocity shifts 
$v_{\rm obs}$ in a fraction of the type 1 quasars.
Bonning et al. (2007; B07 hereafter) set limits on the fraction of emission line
velocity shifts observed in a sample of SDSS quasars.
In our model, the majority ($\gtrsim 75\%$) of kicked SMBHs remain
bound to the galaxy and so their velocities quickly drop
{\em below the initial value of $\vk$}; most of the time
they would therefore be observed with a velocity that is
much smaller than $\vk$.

We carried out Monte-Carlo simulations to check the 
consistency of our recoil model with the B07 limits.
Our simulations were designed to crudely mimic the properties 
of a sample of quasars selected to exhibit both broad and narrow
emission lines, as in the B07 sample.
We adopted a uniform logarithmic distribution of galaxy masses,
$11\le \log_{\rm 10}(M_{\rm gal}/\msun)\le 12.5$, and a fixed
quasar lifetime of $t_{\rm qso}=3\times 10^7$ yr; kicks were assumed
to have occurred at times that were distributed uniformly and randomly between
the epoch of observation and a time $t_{\rm qso}$ earlier.
The position and velocity of the SMBH at the time of observation
was extracted from the appropriate $N$-body model 
after scaling to physical units using the galaxy mass.
If the distance of the SMBH from the galaxy center exceeded the outer 
torus radius, its radial velocity $v_{\rm obs}$
was added to the cumulative distribution  (assuming a random
direction for the recoil);
this velocity was identified with the measured velocity offset 
of the BLR gas in the B07 galaxies.
A fraction 30\% of SMBHs (the ``true'' type 1 population) with $r<r_{\rm tor,out}$ were 
assumed to have visible BLRs and so were included in the
accounting.
Figure~1c shows that the predicted fraction of objects with large
($\approxgt 100$ km s$^{-1}$) velocity shifts is $\sim$ an
order of magnitude smaller than would be inferred from
the unmodified distribution of kick velocities; this is due to the deceleration
that occurs as the SMBH moves through the galaxy.

B07 found a maximal fraction of $f_{500}$=0.04 
quasars with velocity shifts above $v_{\rm obs}=500$ km s$^{-1}$,
and a fraction $f_{1000}$=0.0035 with shifts above  
$v_{\rm obs}=1000$ km s$^{-1}$.
These limits are consistent with our baseline model for the
kick distribution, particularly if we impose an upper limit
to the galaxy/SMBH separation at which a recoiling SMBH
would be spectroscopically identified with its host galaxy
(shown as the dotted lines in Figure~1c, assuming $r_{\rm max}=10$ kpc).

Finally, we note that recoil oscillations will also have a number of other 
observable consequences.
They will affect the X-ray background and its modeling since
a fraction of sources will be unobscured at any given time. 
In particular, small amplitude oscillations of the order the torus size 
will affect the ratio of Compton-thin to Compton-thick sources, and  
could lead to measurable variability in the absorption and extinction 
of AGN spectra once the recoiling SMBH passes the individual clouds 
making up the torus.
A number of interesting effects are related to the torus itself: 
(1) A recoiling SMBH with a bound gas disk that passes through 
the dense torus  (rather than moving perpendicular to it) might 
cause local shocks, heating, and temporary X-ray emission.
(2) During the long-lived ``Phase II'' oscillations, 
when the SMBH oscillation amplitude is on the torus scale, 
the SMBH might efficiently accrete from the dense molecular gas
at {\em each} turning point, causing repeated flares of radiation.
Such flares would locally destroy the dust, while photoionization of 
the dense surrounding gas would produce a strong emission-line response
which would not only help in identifying recoils but could also be used 
as a new probe of the properties of the torus itself.
{\footnote{While strong emission-line variability
in response to an X-ray flare has
recently been observed, this event is more likely interpreted in terms of
stellar tidal disruption (Komossa et al. 2008b).}   
(3) Torus radii are roughly equal to SMBH gravitational
influence radii, so ejection of the SMBH
might lead to temporary expansion of the torus 
since the mass holding it in place is suddenly removed.
Isotropic expansion would not affect the column density along 
the line of sight, seen from the very center.
These effects will be addressed in more detail in
forthcoming work.  

In summary, we have shown that timescales of recoil oscillations
are in an interesting regime where they can potentially
affect a significant fraction of the quasar population.
Details of the predictions still depend on uncertainties in the observed
numbers of AGN types,
quasar lifetimes, torus properties,
structural parameters
of luminous quasar host galaxies, and SMBH spin distributions on the one hand,
and on modelling recoil in realistic non-spherical or non-axially-symmetric galaxies
on the other hand.  
Knowledge of recoil oscillation timescales and amplitudes
is also critical for modelling AGN evolution, delays between starburst and AGN
activity after merger, 
and the cosmic X-ray background. 

\acknowledgments
DM was supported by grants AST-0807910 (NSF) and NNX07AH15G (NASA).
We thank A. Gualandris for assistance in extracting data from
the $N$-body simulations.


\begin{figure*}
\includegraphics[angle=-90.,width=1.\textwidth]{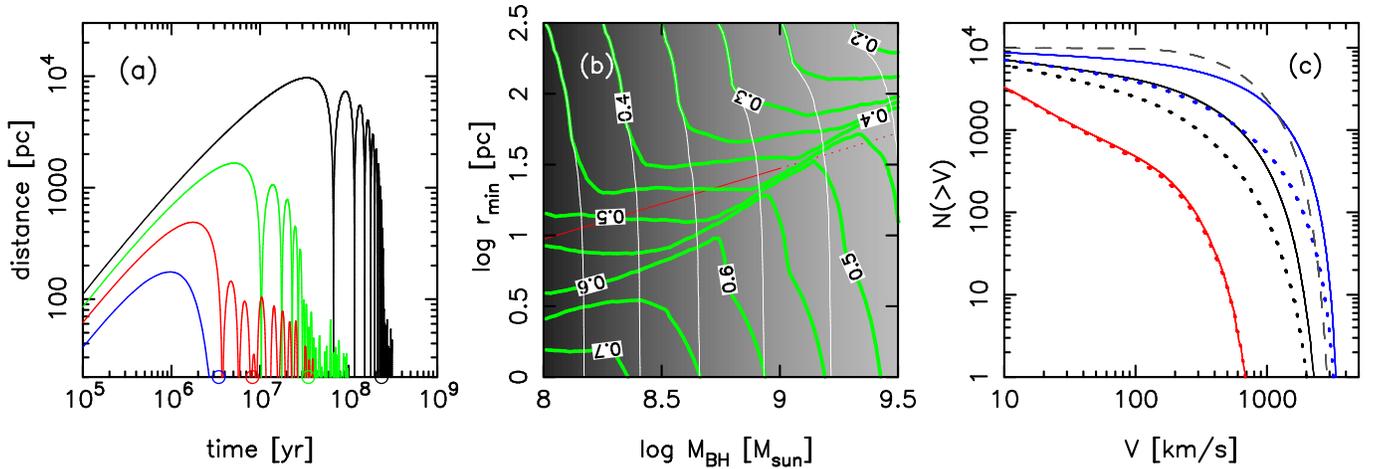}
\caption{
(a) Trajectories of kicked SMBHs in $N$-body models of a
galaxy (based on GM08) with a SMBH mass of
$M_{\rm BH}=5\times 10^{8} M_\odot$, galaxy mass $M_{\rm gal}=5\times 10^{11} M_\odot$, and effective
radius of $7$ kpc. Kick velocities were $(360, 590, 830, 1070)$ km s$^{-1}$
(blue, red, green, black).
Open circles indicate the approximate time at which the
SMBH would come to rest  after ``Phase I'' (see text).
(b) Fraction of kicked SMBHs, in a galaxy with SMBH mass $M_{\rm BH}$, that
remain above the torus, of radius $r_{\rm min}$, for a time of
$3\times 10^7$ yr or longer. Thick
(green) contours are based on the full $N$-body trajectories; thin
(white) contours include only ``Phase I.''  Gray-scale density is proportional
to the fraction of kicks that result in escape, from $\sim 0.3$ on the
left to $\sim 0.05$ on the right.  Kicks were generated assuming
``log-uniform'' distributions of SMBH masses and spins (see text) with
$0.3\le m_2/m_1 \le 1$ and $0.5\le a_{1,2}\le 0.9$.  The red line shows
the approximate outer radius expected for the obscuring torus (based on
eqn (3), and converting $M_{\rm BH}$ to $L$ assuming accretion at 0.1$L_{\rm edd}$).
(c) Velocity distributions that would be observed in a
representative sample of quasars, assuming that kicks occur
randomly in time; additional features of the model are described
in the text. The black line uses the baseline model of SMBH masses
and spins in the pre-recoil binary (the same model  used in Fig.~1b).
Red line: like the base model, but $a=0.3$ for both SMBHs.
Blue line: equal-mass SMBHs with maximal spins.
The dotted lines exclude SMBHs that are more than $10$ kpc from the
galaxy center at the moment of observation.
The gray dashed line shows the input kick distribution for the baseline
model, unmodified by motion through the galaxy.}
\end{figure*}

\end{document}